\documentclass{pasj01}
\usepackage{url}
\usepackage{natbib}
\usepackage{booktabs}
\usepackage[switch,mathlines]{lineno}
\usepackage{xcolor}
\usepackage[switch,mathlines]{lineno}
\Received{$\langle$reception date$\rangle$}
\Accepted{$\langle$acception date$\rangle$}
\Published{$\langle$publication date$\rangle$}

\begin{document}

\title{Efficient Identification of Broad Absorption Line Quasars using Dimensionality Reduction and Machine Learning}

\author{Wei-Bo \textsc{Kao}\altaffilmark{1,2,3}}
\email{weikao@student.ethz.ch}

\author{Yanxia \textsc{Zhang}\altaffilmark{1}}
\email{zyx@bao.ac.cn}

\author{Xue-Bing \textsc{Wu}\altaffilmark{2,4}}

\altaffiltext{1}{CAS Key Laboratory of Optical Astronomy, National Astronomical Observatories, Chinese Academy of Sciences, Beijing, 100101, China}
\altaffiltext{2}{Department of Astronomy, School of Physics, Peking University, Beijing, 100871, China}
\altaffiltext{3}{Department of Physics, ETH Zürich, Wolfgang-Pauli-Strasse 27, CH-8093 Zurich, Switzerland}
\altaffiltext{4}{Kavli Institute for Astronomy and Astrophysics, Peking University, Beijing, 100871, China}

\KeyWords{(galaxies:) quasars: absorption lines --- techniques: spectroscopic --- methods: statistical --- methods: data analysis}

\maketitle

\begin{abstract}
Broad Absorption Line Quasars (BALQSOs) represent a significant phenomenon in the realm of quasar astronomy, displaying distinct blue-shifted broad absorption lines. These enigmatic objects serve as invaluable probes for unraveling the intricate structure and evolution of quasars, shedding light on the profound influence exerted by supermassive black holes on galaxy formation. The proliferation of large-scale spectroscopic surveys such as LAMOST (the Large Sky Area Multi-Object Fiber Spectroscopic Telescope), SDSS (the Sloan Digital Sky Survey), and DESI (the Dark Energy Spectroscopic Instrument) has exponentially expanded the repository of quasar spectra at our disposal. In this study, we present an innovative approach to streamline the identification of BALQSOs, leveraging the power of dimensionality reduction and machine learning algorithms. Our dataset is meticulously curated from the SDSS Data Release 16 (DR16), amalgamating quasar spectra with classification labels sourced from the DR16Q quasar catalog. We employ a diverse array of dimensionality reduction techniques, including Principal Component Analysis (PCA), t-Distributed Stochastic Neighbor Embedding (t-SNE), Locally Linear Embedding (LLE), and Isometric Mapping (ISOMAP), to distill the essence of the original spectral data. The resultant low-dimensional representations serve as inputs for a suite of machine learning classifiers, including the robust XGBoost and Random Forest models. Through rigorous experimentation, we unveil PCA as the most effective dimensionality reduction methodology, adeptly navigating the intricate balance between dimensionality reduction and preservation of vital spectral information. Notably, the synergistic fusion of PCA with the XGBoost classifier emerges as the pinnacle of efficacy in the BALQSO classification endeavor, boasting impressive accuracy rates of 97.60\% by 10-cross validation and 96.92\% on the outer test sample. This study not only introduces a novel machine learning-based paradigm for quasar classification but also offers invaluable insights transferrable to a myriad of spectral classification challenges pervasive in the realm of astronomy.
\end{abstract}

\section{Introduction}\label{sec:intro}
It has been well-established that the vast majority of massive galaxies harbor supermassive black holes (SMBHs) at their cores \citep{Kormendy1995}. However, only a small fraction, estimated at 1-5\%, of these SMBHs exhibit luminous and vigorous activity, giving rise to what is commonly known as Active Galactic Nuclei (AGN; \citealt{Yuk2022}). The formidable energy output observed in AGN is fueled by the relentless accretion of surrounding matter onto the SMBH \citep{Antonucci1993, Heckman2014}.

Quasars, or quasi-stellar objects (QSOs), constitute a small yet extraordinary subset of active galactic nuclei (AGN), distinguished by their prodigious luminosity and high-energy characteristics. The seminal discovery of quasars, attributed to Maarten Schmidt in 1963, marked a watershed moment in the annals of astronomy \citep{Schmidt1963}. The comprehensive study of quasars has engendered profound insights into various facets of astrophysics. Noteworthy contributions include elucidating the intricate co-evolutionary dynamics between supermassive black holes (SMBHs) and their host galaxies \citep{Silk1998, King2003, DiMatteo2005, Kormendy2013}, unveiling the formation mechanisms governing the large-scale structures of galaxies \citep{Eisenstein2011, Dawson2013, Dawson2016}, discerning the presence of interstellar matter and intergalactic medium across diverse redshift regimes \citep{Weymann1981, Hassan2018}, and illuminating pivotal processes such as cosmic reionization and accretion phenomena \citep{Shakura1973, Balbus1998, Lovelace1993, Jiang2019}. The profound impact of quasar research extends far beyond the confines of astrophysical inquiry, transcending disciplinary boundaries to enrich our understanding of the cosmos at large.

Broad Absorption Line Quasars (BALQSOs) represent a distinctive class of celestial entities characterized by striking features. These objects prominently display blue-shifted absorption line troughs, indicative of high-velocity gas outflows emanating from the galactic center along our line of sight. Such outflows can attain velocities exceeding 2000 km/s \citep{Weymann1991}, constituting a substantial feedback mechanism for AGN. The consequential impact of these outflows on galactic star formation rates and dust formation processes has been extensively deliberated \citep{Elvis2002, Bower2006}, implicating BAL quasars in the intricate co-evolutionary dynamics between supermassive black holes and their host galaxies. It is noteworthy that the absorption lines observed in BALQSOs predominantly arise from highly ionized atoms, including C IV, Si IV, N V, and O IV, among others \citep{Turnshek1988}. This distinctive spectral signature serves as a hallmark of BALQSOs, elucidating their pivotal role in probing the intricate astrophysical processes governing galactic-scale phenomena.

The prevalence of BALQSOs within the quasar population exhibits a notable dependence on the observational wavelength. In the ultraviolet and optical spectra, BALQSOs constitute approximately 10-30\% of the population \citep{Reichard2003, Trump2006, Gibson2009}, while in the infrared regime, their representation escalates significantly to around 40\% \citep{Dai2008}. The underlying mechanisms governing this wavelength-dependent trend remain elusive, confounding current astrophysical understanding.
Plausible explanations for this phenomenon abound, including the conjecture that numerous intrinsic BALQSOs evade detection in the optical spectra, or that there exists a nuanced relationship between broad absorption lines and the emitted radiation wavelength of quasars \citep{Bruni2019, Petley2022}. 

An enduring enigma in quasar astronomy pertains to the selective presence of broad absorption lines in only certain quasars. One prevailing hypothesis posits that this phenomenon is contingent upon the observer's viewing angle relative to the AGN, where the manifestation of broad absorption lines hinges upon the orientation from which the AGN's broad emission line region is perceived \citep{Ogle1999, Ghosh2007}. Alternatively, an intriguing conjecture postulates that BALQSOs may represent a transient evolutionary phase in the life cycle of quasars, and their incidence rate could correspond to the duration of this evolutionary stage \citep{Lipari2009, Petley2022}.




As the significance of quasars in astrophysical research continues to grow, the pace of quasar discovery has accelerated exponentially. Large-scale surveys, such as the Sloan Digital Sky Survey (SDSS), and forthcoming initiatives like the Vera C. Rubin Observatory Legacy Survey of Space and Time (LSST) and the Chinese Space Station Telescope (CSST), have ushered in an era of unprecedented data volume, inundating astronomers with copious quasar spectral and image data. The burgeoning dataset presents a formidable challenge to traditional manual classification methods, whose efficacy diminishes in the face of such vast quantities of data. Moreover, manual classification inherently suffers from subjective biases. Hence, there is an imperative to develop automated, objective methodologies for quasar classification. For instance, \cite{Paris2017} undertook the laborious task of manually classifying 297,301 quasars in SDSS DR12, identifying 29,580 as BALQSOs. However, the subsequent release of SDSS DR16 \citep{Ahumada2020, Lyke2020} witnessed a meteoric surge in the number of quasars to 750,414, rendering manual classification impracticable. Consequently, automated detection methods have emerged as indispensable tools in quasar classification. These automated methodologies leverage diverse techniques such as Principal Component Analysis (PCA), quasar spectral templates, and metrics like the balnicity index (BI) and absorption index (AI) to estimate the likelihood of a celestial object being a BALQSO \citep{Weymann1991, Hall2002, Guo2019}. By harnessing the power of computational algorithms, these approaches offer an efficient, objective means of navigating the labyrinthine landscape of quasar classification, facilitating insights into the cosmos on an unprecedented scale.

Machine learning has emerged as a transformative tool in astronomy, yielding remarkable successes in diverse applications. For instance, \cite{Fu2021} demonstrated the efficacy of transfer learning coupled with the XGBoost algorithm in identifying quasars obscured by the Galactic plane, leveraging analysis of their multiband photometric data. \cite{Hausen2020} pioneered the use of convolutional neural networks (CNNs) and deep learning techniques to detect and categorize astronomical objects into various classes, including elliptical galaxies, disk galaxies, irregular galaxies, point sources, and background objects. Building on this momentum, \cite{Busca2018} harnessed the power of CNNs for spectroscopic redshift estimation and quasar classification, concurrently exploring the presence of broad absorption line structures. Moreover, \cite{CarrascoKind2014} introduced unsupervised machine learning methodologies, employing self-organizing maps (SOMs) to estimate photometric redshifts of galaxies, exemplifying the versatility of machine learning techniques across diverse astronomical domains. 

In this study, we leverage dimensionality reduction and machine learning algorithms for the classification of BALQSOs in the SDSS Data Release 16 (SDSS DR16). In Section \ref{sec:data}, we provide a comprehensive overview of the dataset employed in our analysis, elucidating the data preprocessing steps and the intricacies of dataset construction. Section \ref{sec:Dimensionality_Reduction} delves into the dimensionality reduction techniques utilized in our study, encompassing Principal Component Analysis (PCA), t-Distributed Stochastic Neighbor Embedding (t-SNE), and manifold learning methods. In Section \ref{sec:ML}, we present the methodology underpinning the introduction, training, and evaluation of our machine learning classifiers, delineating the steps involved in model selection and performance assessment. The classification outcomes derived from each model are discussed in Section \ref{sec:classification result}, where we meticulously examine and interpret the efficacy of our classification framework. Section \ref{sec:discussion} juxtaposes our findings with prior literature, facilitating a nuanced comparative analysis and offering insights into the advancements achieved through our approach. Lastly, Section \ref{sec: conclusion} encapsulates the key findings of our study, elucidating the broader implications of our research and outlining avenues for future exploration.

\section{Data} \label{sec:data}

\subsection{Dataset Construction}
The Sloan Digital Sky Survey (SDSS; \citealt{York2000}) stands as a cornerstone of modern astronomy, yielding an unparalleled bounty of scientific discoveries. By harnessing the 2.5 m Sloan Foundation Telescope at Apache Point Observatory (APO) and the du Pont 2.5 m Telescope at Las Campanas Observatory (LCO), SDSS has facilitated deep multi-color imaging across one-third of the celestial sphere, while also furnishing spectroscopic data for over three million astronomical objects. Since its inception in 2000, SDSS has undergone successive phases of survey operations, each marked by remarkable advancements and scientific achievements. The evolutionary trajectory of SDSS encompasses several distinct phases: SDSS-I (2000-2005), SDSS-II (2005-2008), SDSS-III (2008-2014), SDSS-IV (2014-2020), and the ongoing SDSS-V initiative (2020–2025). Notably, SDSS-IV comprises three pivotal surveys: the Extended Baryon Oscillation Spectroscopic Survey (eBOSS; \citealt{Dawson2016}), the APO Galactic Evolution Experiment 2 (APOGEE-2; \citealt{Majewski2017}), and Mapping Nearby Galaxies at APO (MaNGA; \citealt{Bundy2015}). 

The spectral data utilized in this study originates from the 16th Data Release of the Sloan Digital Sky Survey-IV (SDSS DR16; \citealt{Ahumada2020}), representing a significant milestone in the ongoing SDSS-IV initiative. SDSS DR16 serves as the fourth major data release within the SDSS-IV project, offering a comprehensive repository of astronomical spectra. The spectra archived within SDSS-IV/eBOSS are captured using 500 fibers on a 2k CCD, spanning a wavelength range of 3600–10400 {\AA}, with a spectral resolution of approximately 2000. This expansive dataset empowers astronomers with a wealth of spectroscopic information, enabling detailed analyses of celestial objects spanning diverse astrophysical regimes.

The DR16 quasar catalog (DR16Q; \citealt{Lyke2020}) includes 920,110 observations of 750,414 quasars. DR16Q is the final quasar catalog of the SDSS-IV/eBOSS and comprises the largest sample of spectroscopically confirmed quasars available to date. 

A plethora of derived quantities extracted from the quasar spectra within the SDSS DR16 are meticulously cataloged in the DR16Q database. Among these derived quantities, redshift estimates hold paramount significance, serving as fundamental inputs for numerous scientific investigations spanning diverse astronomical domains. Notably, redshift estimates play a pivotal role in elucidating the large-scale distribution of cosmic structures, probing the subtle signatures of baryon acoustic oscillations (BAOs), and unraveling the intrinsic properties of quasars themselves \citep{Hou2021,duMasDes2020}. To mitigate inherent limitations and systematic errors associated with different redshift estimation methodologies, DR16Q implements a diverse array of approaches. These approaches are meticulously curated and rigorously validated to ensure the robustness and reliability of the redshift estimates. By leveraging a multifaceted strategy encompassing various estimation techniques, DR16Q endeavors to furnish astronomers with a comprehensive and meticulously vetted dataset, fostering a deeper understanding of the cosmos and facilitating groundbreaking scientific discoveries. These approaches are listed below. 

\begin{enumerate}
    \item The SDSS pipeline. It fits various spectral templates and estimates the redshift based on the one with the smallest fitting error \citep{Bolton2012}.
    \item The QuaserNET algorithm. It utilizes deep CNNs to estimate redshifts from spectra \citep{Busca2018}. 
    \item Manual estimation. Some quasars' redshifts are visually inspected in DR16Q.
    \item Estimated by principal component analysis (PCA). Fitting the first four principal components of the spectra along with a quadratic polynomial to estimate the redshifts.
\end{enumerate}

In DR16Q, a methodology for selecting a ``primary" redshift for each quasar is outlined. This process prioritizes manual inspection redshifts whenever high-confidence inspection is available, resorting to SDSS automated pipeline redshifts otherwise. The primary redshift is documented in the column ``Z". Since it is expected to offer a least-biased estimate of quasar redshifts, we use this primary redshift for the analysis in this study.

In the DR16Q, the probability of a quasar exhibiting BAL features (BAL\_PROB) with $1.57 \leq z \leq 5.6$ is also estimated \citep{Lyke2020}. The procedure is introduced here. Firstly, the $\chi^2$ fit of unabsorbed quasar models is performed to each quasar spectrum. Then, the identified regions with obvious differences between the best-fit spectrum and the original spectrum would represent absorption. These regions are masked. They iterated these steps of fitting, identifying absorption region, and masking. The absorption features were then characterized by the balnicity index (BI), absorption index (AI), and their uncertainties ($\sigma_{\textrm{BI}}$ and $\sigma_{\textrm{AI}}$). BAL\_PROB is computed based on $\chi^2$ uncertainties from the spectrum fitting and BI and AI from C IV absorption features. The resulted values of BAL\_PROB are discontinuous with discrete values of 0, 0.5, 0.75, 0.9, 0.95, or 1.

\subsection{Data Preprocessing} \label{subsec:preprocess}
Data preprocessing is a crucial step in improving the accuracy of a model by transforming raw data into a clean and well-formatted dataset. Each raw galaxy spectrum in SDSS consists of flux data points that are distributed logarithmically with respect to wavelength. Initially, we extracted the wavelength and corresponding flux from each spectrum. Additionally, spectra that have ZWARNING flag in DR16Q are excluded from the dataset.

Subsequently, we utilized the primary redshift in DR16Q to transform each spectrum into rest-frame wavelength space. We chose a rest-frame wavelength range of 1260-2400 {\AA} to study the structure of spectral lines such as C IV and Si IV, as well as potential absorption. This wavelength range provides comprehensive coverage of the C IV and Si IV regions, enabling us to characterize C IV absorption troughs with blueshifts up to 25,000 km/s while minimizing the investigation of unnecessary regions. Some observed spectra do not encompass this entire rest-frame wavelength range and were consequently excluded from our analysis.

For processing the spectral data within this wavelength range, we employed linear interpolation with an interval of 0.8 {\AA}. This method resulted in all spectra having 1424 data points and reduced the impact of errors. To normalize the flux of each spectrum, we scaled the spectrum flux by dividing it by the average flux of that spectrum, such that the mean flux of each spectrum is normalized to 1. To clarify, we denoted the flux of one spectrum as an $m$-dimensional vector $x = (x_1,x_2,...,x_m)$ and normalized the flux by employing the following equation:
\begin{equation}
    x' = \frac{x}{\sum_{i=1}^m \frac{x_i}{m}}.
\end{equation}

\begin{figure*}
\centering
\includegraphics[width=0.8\linewidth]{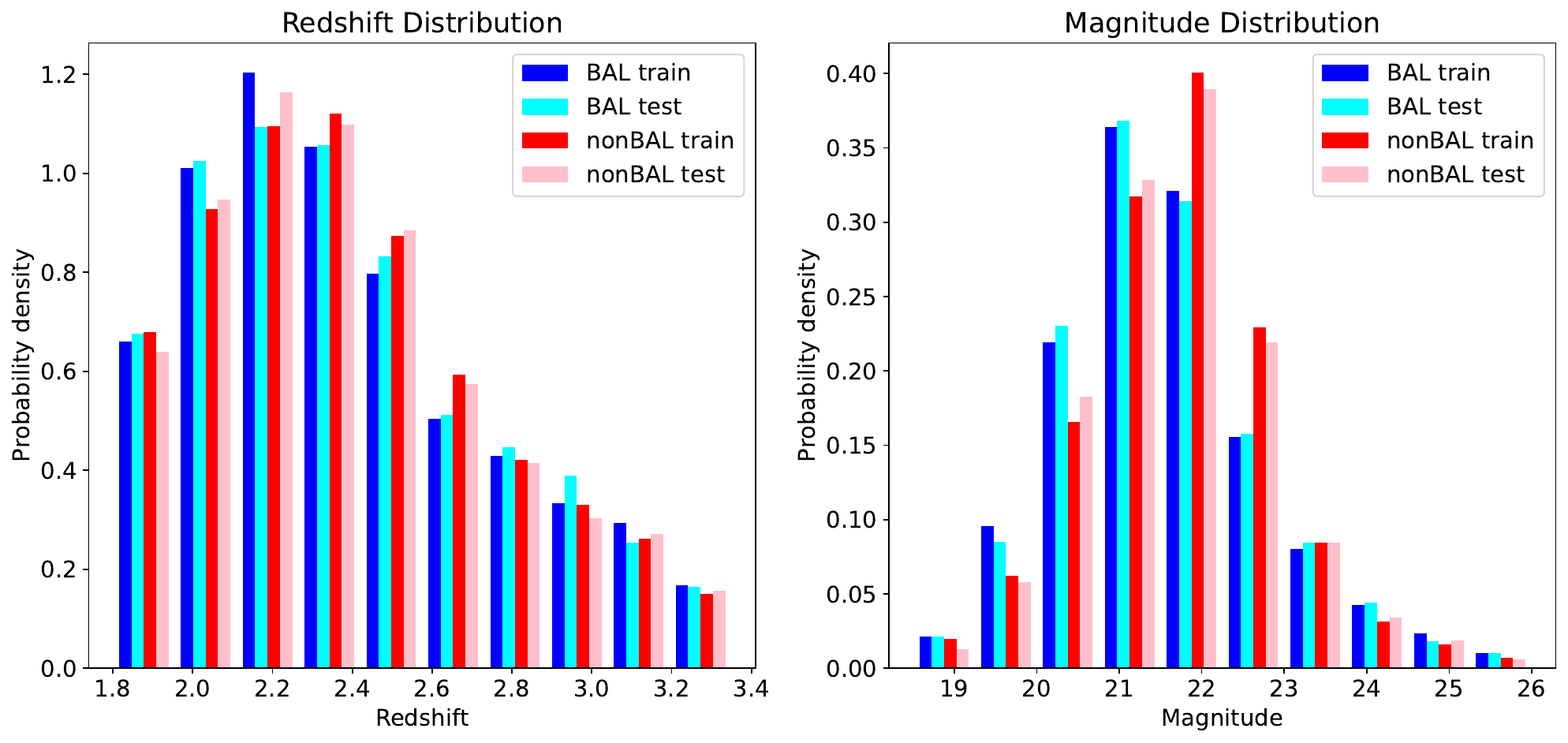}   
\caption{Distribution of redshifts and magnitudes for BALQSOs in the training (blue) and test (cyan) samples, as well as non-BALQSOs in the training (red) and test (pink) samples.
\label{fig:train_test}}
\end{figure*}

In our study, we define BALQSOs and non-BAL QSOs based on the BAL\_PROB parameter in the DR16Q catalog. We adopt a strict selection criterion, where objects with BAL\_PROB $=0$ are classified as non-BAL QSOs, while those with BAL\_PROB $=1$ are labeled as BALQSOs. After data preprocessing, we have 313,735 non-BAL QSOs and 15,733 BALQSOs by this definition. For balancing the data, we randomly selected 15,733 non-BAL QSOs and all BALQSOs to construct the dataset for training and testing the model. Additionally, 10\% of the data was randomly selected to construct a outer test sample for blind testing of the model, comprising 1574 non-BAL QSOs and 1574 BALQSOs. The remaining data were used to form the training sample, which comprises 14,159 non-BAL QSOs and 14,159 BALQSOs. Figure \ref{fig:train_test} illustrates the redshift and PSF magnitude in $u$ band distribution for BALQSOs and non-BAL QSOs in both training and test samples. The disparity between the distribution of training and test samples is small. 


\section{Dimensionality Reduction} \label{sec:Dimensionality_Reduction}

High-dimensional data are prevalent in astronomy and machine learning, such as spectral data and image data. These raw data contain rich information but often include redundant or unnecessary information that is irrelevant or even detrimental to data analysis and processing. Dimensionality reduction techniques aim to transform high-dimensional data into a lower-dimensional representation. By doing so, it becomes easier to understand and visualize the data. Data dimensionality can also reduce the data volume and complexity, saving computational time. Additionally, it can mitigate the issues of overfitting and noise, thereby enhancing the efficiency and accuracy of model training and prediction.

In our study, we first preprocess the data and then explore common dimensionality reduction techniques such as principal component analysis (PCA), t-SNE (t-distributed Stochastic Neighbor Embedding), and manifold learning algorithms. We apply these techniques to reduce the dimensionality of our data. All dimensionality reduction methods are implemented using the scikit-learn 1.2 library in Python. The resulting lower-dimensional representation is then used as input for machine learning algorithms to classify quasars.

\subsection{Principal Component Analysis}\label{subsec:pca}
Principal Component Analysis (PCA) stands as a ubiquitous data dimensionality reduction technique widely employed in astronomical spectral analysis. PCA plays a pivotal role in distilling essential information from high-dimensional datasets, thereby simplifying the data structure and expediting subsequent analytical procedures.

Through linear transformations, PCA adeptly maps the high-dimensional dataset onto a lower-dimensional subspace while endeavoring to preserve the maximum variance inherent in the original data. This transformation yields a set of new low-dimensional features termed principal components. Notably, these principal components are ordered such that the first principal component encapsulates the highest variance, followed by successive components in descending order of variance.

A distinctive advantage of PCA lies in its parameter-free nature, facilitating straightforward implementation and interpretation. This attribute renders PCA particularly appealing for various spectral data analyses in astronomy, spanning classification tasks, variability analysis, denoising procedures, and myriad other applications \citep{VandenBerk2006, Zhang2006, Bailey2012}.

By leveraging PCA's innate capabilities, astronomers can effectively distill critical insights from complex spectral datasets, thereby advancing our understanding of celestial phenomena and enabling groundbreaking discoveries in the field of astronomy.

Initially, we conducted PCA for dimensionality reduction. The dataset utilized for this analysis was the test data obtained following the data preprocessing steps outlined in Section \ref{subsec:preprocess}, comprising 14,159 non-BAL QSOs and 14,159 BALQSOs. PCA was performed on the entirety of the spectral data, retaining solely the first 100 principal components derived from the PCA transformation.

The variances associated with the first 8 principal components, also known as eigenvectors, within the entire dataset were determined to be 7.3\%, 5.1\%, 4.4\%, 3.7\%, 3.2\%, 2.8\%, 2.7\%, and 1.4\%, respectively. These variances serve as indicators of the proportions of information retained by each principal component. Notably, the cumulative variance attributed to these 8 principal components amounts to 30.7\%. 

The cumulative variance of the first 100 principal components in the large sample is 58\%, indicating that these components retain 58\% of the information. This reduction in dimensionality from 1424 dimensions to 100 dimensions corresponds to a reduction of 93\% in the number of dimensions, demonstrating the effectiveness of the dimensionality reduction technique.

\subsection{t-SNE} \label{subsec:tsne}
t-SNE (t-distributed stochastic neighbor embedding) serves as a nonlinear dimensionality reduction technique designed to project high-dimensional data into a more visually interpretable two- or three-dimensional space. Initially introduced by Laurens van der Maaten and Geoffrey Hinton in 2008, t-SNE represents an advancement over the earlier SNE algorithm \citep{vanDerMaaten2008}.

In essence, t-SNE operates by first computing the pairwise similarity between each pair of samples within the high-dimensional space. It then transforms these similarities into a probability distribution, wherein close samples are assigned higher probabilities while distant samples receive lower probabilities. Crucially, t-SNE employs the t-distribution to define the probability distribution in the low-dimensional space, assigning higher probabilities to nearby samples and lower probabilities to those farther apart.

During the transformation process, t-SNE endeavors to minimize the Kullback-Leibler (KL) divergence between the probability distributions in the high-dimensional and low-dimensional spaces via gradient descent. By doing so, t-SNE aims to maximize the preservation of local relationships between samples within the high-dimensional space, thereby facilitating a faithful representation of the intrinsic structure of the data in the lower-dimensional embedding.

The primary advantage of t-SNE lies in its capability to effectively visualize category information and accentuate the local structure inherent in high-dimensional datasets. In contrast to PCA, t-SNE exhibits stronger nonlinear capabilities and provides enhanced interpretability, allowing for more intuitive insights into the underlying data structure. However, it is important to note that t-SNE entails a higher computational cost and is subject to certain limitations.

One such limitation is t-SNE's sensitivity to initial randomization and the presence of noise or outliers within high-dimensional data. Consequently, careful preprocessing and outlier detection strategies are essential to mitigate these potential challenges. Furthermore, the selection of appropriate parameters, such as perplexity, significantly influences the effectiveness of t-SNE dimensionality reduction. Ensuring optimal parameter settings is crucial for obtaining meaningful and reliable results when employing t-SNE in data analysis tasks.

Despite these limitations, t-SNE remains a valuable tool for exploratory data analysis and visualization, particularly in scenarios where capturing complex local relationships within the data is paramount. By leveraging its nonlinear capabilities and interpretative prowess, researchers can gain deeper insights into the underlying structure of high-dimensional datasets, ultimately facilitating more informed decision-making and hypothesis generation in various scientific domains, including astronomy, computer vision, natural language processing, and social network analysis. In astronomy, t-SNE has been extensively used for tasks such as spectral classification, stellar classification, and identification of anomalous spectral data, enabling astronomers to gain a better understanding of the intrinsic structure of high-dimensional astronomical data \citep{Verma2021, Anders2018, Traven2017}.

\begin{figure}
\centering
\includegraphics[width=0.75\linewidth]{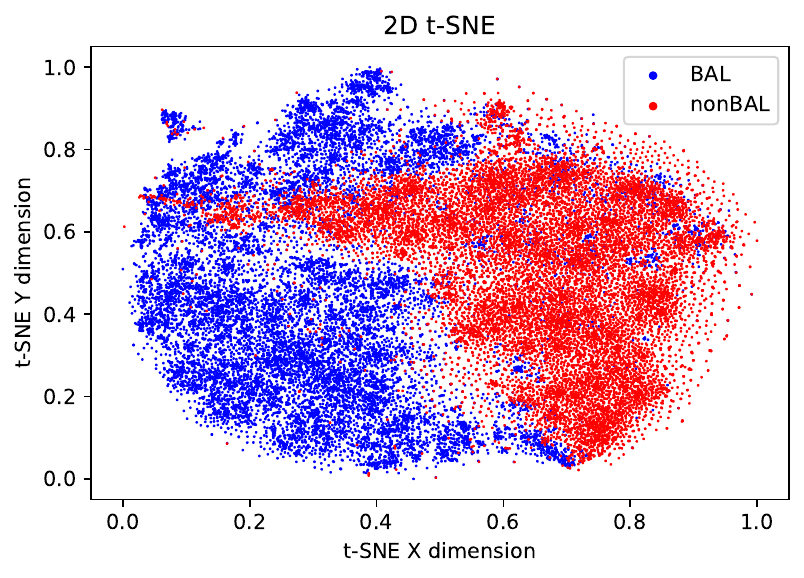}    
\caption{Two-dimensional representation of dataset reduced by t-SNE. t-SNE is applied to reduce the dimensionality of non-BAL QSOs (red points) and  BALQSOs (blue points) to two dimensions. 
\label{fig:tSNE_2D}}
\end{figure}

In this investigation, we employed t-SNE dimensionality reduction on the train dataset (outlined in Section \ref{subsec:preprocess}). Initially, we directly fed all data points from the dataset into t-SNE with default parameters, aiming to reduce the dimensionality to two dimensions. The resulting visualization, depicted in Figure \ref{fig:tSNE_2D}, illustrates t-SNE's ability to roughly segregate the two classes of QSOs. However, it's noteworthy that the boundary between these classes isn't a straightforward linear separation and isn't readily discernible. Consequently, for precise classification of the two QSO types, a combination with machine learning techniques becomes imperative.

This finding underscores the complementary nature of t-SNE and machine learning methods, where t-SNE facilitates the initial exploration and understanding of the data's intrinsic structure, while machine learning algorithms further refine and categorize the data based on these learned patterns. By leveraging both t-SNE for dimensionality reduction and machine learning for classification, we can achieve a comprehensive and accurate analysis of the dataset, ultimately enhancing our understanding of the underlying properties and characteristics of the observed celestial objects.

To evaluate the influence of various dimensionality reduction methods and parameters on the efficacy of the machine learning classifier, we investigated the following three t-SNE dimensionality reduction strategies. Subsequently, we leveraged the outcomes of these approaches as input data for the machine learning models:

\begin{enumerate}
    \item Apply t-SNE directly: We directly inputted all the data points from the dataset into the t-SNE algorithm, and reduced the high-dimensional data to two-dimensional and three-dimensional using the default parameters.
    \item Reduce the data by PCA then apply t-SNE: We first performed PCA on the dataset, reducing it to 100 dimensions. Subsequently, we inputted the reduced data into the t-SNE algorithm, obtaining two-dimensional and three-dimensional representations using the default parameters.
    \item Change the ``perplexity" parameter of t-SNE: We inputted all data from the dataset into the t-SNE algorithm, adjusting the perplexity parameter to different values. Perplexity determines the number of nearest neighbors used in the algorithm and is generally chosen based on the dataset size. The default value is 30, and we experimented with perplexity values of 5, 10, 30, 50, and 100.
\end{enumerate}

By employing these different dimensionality reduction methods and parameter combinations, we obtained multiple sets of low-dimensional representations for subsequent training and testing of the machine learning model (See Section \ref{sec:ML}).

\subsection{Manifold Learning} \label{subsec:manifold_learning}

Manifold learning encompasses a collection of nonlinear dimensionality reduction techniques aimed at uncovering and exploring the intrinsic structure or geometry of high-dimensional datasets. It operates under the assumption that the data points are distributed on or near a lower-dimensional manifold embedded within the higher-dimensional space. The primary objective of manifold learning is to unveil this latent manifold and project the data onto a lower-dimensional subspace while preserving its essential characteristics.

A prototypical example frequently used to demonstrate the efficacy of manifold learning techniques is the Swiss roll dataset, which exists in three-dimensional space. Traditional distance metrics, such as Euclidean distance, may fail to adequately capture the inherent similarity relationships between data points in the Swiss roll due to its curved and twisted nature. In such cases, manifold learning methods come into play, leveraging local relationships and intrinsic manifold structures to more accurately capture the underlying similarity structure of the data. Through this process, manifold learning facilitates a more comprehensive understanding of the dataset's underlying geometry and enables effective visualization and analysis of complex high-dimensional datasets.

Locally Linear Embedding (LLE) and Isometric Mapping (ISOMAP) are prominent manifold learning algorithms commonly employed in dimensionality reduction tasks \citep{Tenenbaum2000, Roweis2001}. Both LLE and ISOMAP aim to capture the local structure of the data and project it into a lower-dimensional space, effectively revealing the underlying manifold structure of the dataset.

The LLE algorithm operates by preserving the local linear relationships within the data during the dimensionality reduction process. It achieves this by iteratively reconstructing each data point as a linear combination of its nearest neighbors in the high-dimensional space. By preserving these local linear relationships, LLE effectively maps the data into a lower-dimensional space while retaining the original data's intrinsic structure.

In contrast, ISOMAP focuses on capturing the manifold's global geometry by estimating the geodesic distances between data points, representing distances along the manifold's curved surface. It then employs multidimensional scaling (MDS) to embed the data points into a lower-dimensional space based on these geodesic distances. This approach allows ISOMAP to accurately preserve the underlying manifold's shape and structure, even in cases where the data exhibits nonlinear relationships.

In the field of astronomy, both LLE and ISOMAP have found widespread application in tasks such as dimensionality reduction, visualization, and classification of galaxy and stellar spectroscopic data \citep{Bu2014, Daniel2011, Matijevic2012}. These algorithms offer astronomers valuable tools for exploring and interpreting complex astronomical datasets, providing insights into the underlying structures and relationships within the data. From a scientific perspective, the application of LLE and ISOMAP in astronomy contributes to a deeper understanding of celestial phenomena and facilitates discoveries in the field.

\begin{figure*}
\centering
\includegraphics[width=0.8\linewidth]{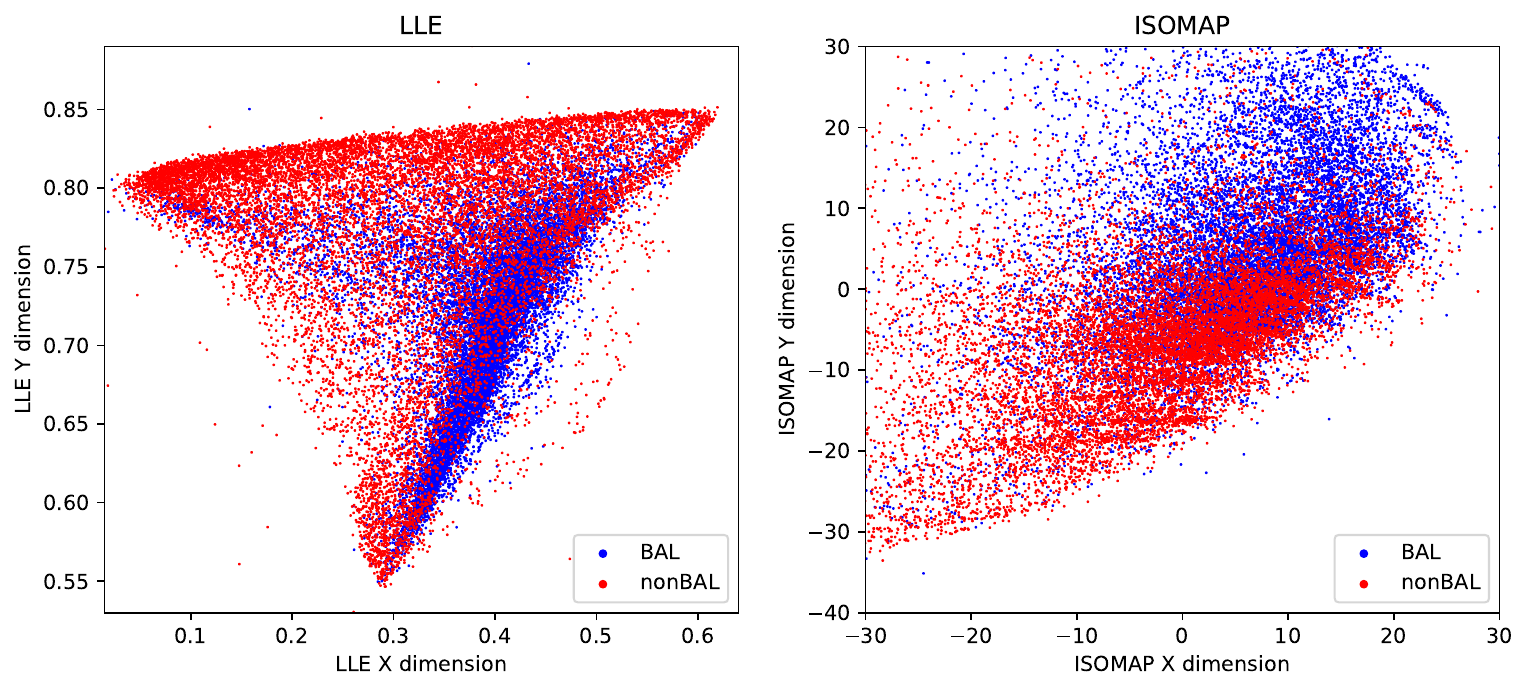}    
\caption{Two-dimensional representation of data points reduced by the LLE and ISOMAP algorithms. The dataset contains 14,159 non-BAL QSOs (red points) and 14,159 BALQSOs (blue points). It is clear that both manifold learning algorithms cannot effectively separate two classes of QSOs.
\label{fig:LLE_ISOMAP}}
\end{figure*}

We also employed the ISOMAP and LLE dimensionality reduction algorithms on the training dataset to conduct nonlinear dimensionality reduction analysis. Initially, we inputted all the data into these algorithms and reduced the dimensionality to two dimensions. As shown in Figure \ref{fig:LLE_ISOMAP}, the visualizations of the reduced data from both methods did not effectively segregate the two classes of quasars in the two-dimensional space. Consequently, dimensionality reduction of the data needs to be further considered as the input of a machine learning model for classification.

To address this, we subsequently reduced the dimensionality of the data using ISOMAP and LLE to 10 and 100 dimensions, respectively. The resulting embeddings were then utilized as the input for machine learning models to perform classification (See Section \ref{sec:ML}). This approach aimed to enhance the separation of the classes and improve the performance of the subsequent classification task. From a scientific standpoint, this iterative process of dimensionality reduction and classification ensures that the machine learning models can effectively capture the underlying structure of the data and accurately classify the quasar subclasses. Through this methodology, we strive to uncover meaningful insights from the astronomical data and contribute to the advancement of knowledge in the field.

\section{Machine Learning Algorithms for Classification} \label{sec:ML}

In Section \ref{sec:Dimensionality_Reduction}, we explored various dimensionality reduction methods and obtained corresponding results. After reducing the dimensionality, we utilized the reduced data as inputs to train different machine learning models and evaluate their respective accuracies. By comparing the accuracies achieved under different scenarios, we can assess the performance of different combinations of dimensionality reduction methods and machine learning models for the classification task. This analysis helps us discern the strengths and limitations of each method, aiding in the selection of the most suitable combination for our specific task. Through these experiments, we can draw conclusions and offer valuable insights for future similar classification tasks. 

\subsection{Introduction to the Machine Learning Algorithms} \label{subsec: intro_ML}

In this section, we introduce the classification algorithms employed in this study, namely the Random Forest algorithm and the XGBoost algorithm. We elucidate their applications, fundamental principles, and model parameters. In this study, Random Forest is implemented using the scikit-learn 1.2 library in Python, while XGBoost is implemented using the XGBoost 1.7 library in Python.

\subsubsection{Random Forest}

Random Forest \citep{Ho1995, Breiman2001} is a supervised machine learning algorithm commonly employed for classification, regression, and feature selection tasks. It operates as an ensemble learning algorithm built upon decision trees, harnessing the benefits of ensemble learning to attain high accuracy and stability. Random Forest fabricates multiple decision trees by randomly selecting features and samples (bootstrapping), subsequently amalgamating them to make predictions. The final prediction is derived through averaging the predictions from all decision trees or employing majority voting.

This algorithm integrates two primary sources of randomness: feature randomization and sample randomization. During the construction of each decision tree, it randomly cherry-picks a subset of features from all available features and constructs the decision tree based on these selected features. This strategy helps alleviate overfitting concerns associated with individual deep trees. As different trees utilize distinct subsets of features, it can enhance the model's generalization capability. Regarding sample randomization, Random Forest randomly picks a subset of samples from the original dataset to construct each decision tree, thus mitigating the influence of noise and variance in the dataset. Furthermore, each decision tree within the Random Forest ensemble operates independently, facilitating parallel execution and significantly enhancing computational efficiency.

Random Forest finds extensive application in the identification and classification tasks of celestial objects such as galaxies, stars, and planets. For example, it has been utilized for classifying galaxy morphology based on morphological parameters and other features \citep{Rose2023}, as well as for classifying variable stars based on their color, morphology, and additional features. Additionally, it can be employed for estimating photometric redshifts \citep{Carliles2010} and for processing and analyzing astronomical spectroscopic data \citep{Echeverry2022}.

Random Forest has several hyperparameters that can impact its performance. Here are some common and important hyperparameters:
\begin{enumerate}
    \item n\_estimators: This parameter specifies the number of decision trees in Random Forest. Increasing the value of n\_estimators may improve the performance but increases computational cost. The default value is 100.    
    \item max\_depth: This parameter determines the maximum depth allowed for each decision tree. Increasing max\_depth may improve the accuracy but also faces the risk of overfitting.     
    \item max\_features: This parameter specifies the number of features to consider when training each decision tree. Increasing max\_features can improve the performance. However, the risk of overfitting and computational cost becomes higher. 
\end{enumerate}

Through our experimental tuning, we have determined that the n\_estimators hyperparameter holds significant influence over the classification performance of Random Forest. As a result, our primary emphasis is on fine-tuning n\_estimators to optimize the classification performance in subsequent tasks.

\subsubsection{XGBoost}

XGBoost (eXtreme Gradient Boosting) is an efficient machine learning algorithm proposed in 2016 \citep{Chen2016}, building upon the concept of Gradient Boosting Decision Trees \citep{Friedman2001}. Notably, XGBoost exhibits remarkable prowess in handling high-dimensional data and large-scale datasets, making it a favored choice for classification and regression tasks.

In XGBoost, decision trees serve as the foundational classifiers, and the model is trained through gradient boosting, where each iteration introduces a new decision tree to rectify the prediction errors of the preceding trees. This iterative approach significantly enhances model accuracy and facilitates the handling of extensive datasets. Moreover, XGBoost implements feature subsampling akin to Random Forest, randomly selecting a subset of features for constructing each tree. This feature subsampling technique contributes to the model's generalization capabilities by diversifying tree constructions. Coupled with L1/L2 regularization, XGBoost effectively combats overfitting risks. To further enhance computational efficiency, XGBoost adopts strategies like greedy algorithms and parallel computing.

XGBoost has emerged as a formidable tool for classification and prediction tasks in astronomy. It has found applications in classifying galaxies, stars, and quasars based on spectroscopic data or multi-band photometric data \citep{Li2019, Li2021, Jin2019}. Additionally, XGBoost has been instrumental in identifying quasar candidates and detecting red giant clusters \citep{He2022, Fu2021}.

Similar to Random Forest, XGBoost has several important hyperparameters that can be tuned to affect classification results. Here are some commonly used ones:
\begin{enumerate}
    \item max\_depth: It specifies the maximum depth of each tree. A higher value increases the model's complexity but may lead to overfitting. A lower value may result in underfitting. The default value is 6.    
    \item learning\_rate (eta): It controls the step size when updating the weights during each iteration. A smaller learning rate makes the model more stable but requires more iterations to achieve decent results. The default value is 0.3.
\end{enumerate}

Through our rigorous experimentation and fine-tuning processes, we have determined that the max\_depth hyperparameter plays a pivotal role in shaping the classification performance of XGBoost. Consequently, in our subsequent machine learning classifier, we have prioritized the optimization of this hyperparameter to attain optimal results. Our meticulous approach aims to harness the full potential of XGBoost and elevate the accuracy and effectiveness of our classification model.

\subsection{Model Training Method} \label{subsec:training method}

To achieve optimal classification performance, the method employed for training the machine learning model is paramount. In our approach, we adopt cross-validation, specifically 10-fold cross-validation, to enhance accuracy and mitigate the risk of overfitting.

The underlying principle of $k$-fold cross-validation involves partitioning the original dataset into $k$ subsets, typically with $k=5$ or $k=10$. During each iteration, one subset is designated as the test set, utilized to assess the model's accuracy. Meanwhile, the remaining $k-1$ subsets serve as the training set, facilitating model training. This process is repeated $k$ times, with each subset serving as the test set in turn. Subsequently, performance evaluation metrics such as Accuracy, Recall, and F1-Score are computed for each iteration. Finally, the average performance evaluation across all iterations provides a comprehensive assessment of the model's effectiveness.

The adoption of $k$-fold cross-validation offers several advantages, including the utilization of the entire dataset, mitigated risk of overfitting, and enhanced accuracy in evaluating the model's performance. However, it necessitates multiple model training iterations, resulting in increased computational overhead. Despite this, the benefits accrued from the rigorous assessment afforded by $k$-fold cross-validation justify its computational cost.

Our approach to model training is as follows:
\begin{enumerate}
    \item Choose an appropriate dimensionality reduction method based on the discussion in Section \ref{sec:Dimensionality_Reduction}. Input the dimensionality-reduced test data into the selected machine learning model (see Section \ref{subsec: intro_ML}). Also, input the classification labels of the quasars to train the model.
    \item Train the model using the default hyperparameters and perform 10-fold cross-validation to optimize the model in terms of accuracy.
    \item If needed, adjust the hyperparameters of the machine learning model to further enhance the model's performance.
\end{enumerate}

By following this approach, we can systematically train the models using various dimensionality reduction methods and machine learning classifiers. Subsequently, we evaluate each model's performance using 10-fold cross-validation. This rigorous evaluation process enables us to compare the effectiveness of different combinations of methods and hyperparameters for our specific classification task. Through this iterative process of training, evaluation, and refinement, we aim to identify the most effective approach for accurately classifying quasar spectra, ultimately advancing the state-of-the-art in astronomical data analysis.

\section{Classification Result} \label{sec:classification result}

In Section \ref{sec:Dimensionality_Reduction}, we extensively explored different dimensionality reduction techniques on our spectroscopic dataset. Subsequently, we utilized the dimensionality-reduced data as the input for two machine learning models: Random Forest and XGBoost. The labels corresponding to BALQSOs and non-BAL QSOs were utilized as the training targets. To enhance accuracy, we implemented 10-fold cross-validation and adhered to the methodologies delineated in Section \ref{subsec:training method} for model training. This comprehensive approach ensures robust training and evaluation of our models, aiming to achieve superior classification performance.

Accuracy, Precision, Recall and F1-Score are defined by the equations below:
\begin{eqnarray}
    \mathrm{Accuracy} &=& \frac{TP+TN}{TP+FN+FP+TN} \label{eq:accuracy} \\
    \mathrm{Precision} &=& \frac{TP}{TP+FP} \\
    \mathrm{Recall} &=& \frac{TP}{TP+FN}\\
    \mathrm{F1-Score} &=& 2 \times \frac{\rm Precision \times \rm Recall}{\rm Precision + \rm Recall}
\end{eqnarray}
where $TP$ represents the true positive, which refers to correctly classified BALQSOs; $TN$ represents the true negative, which refers to correctly classified non-BAL QSOs; $FP$ represents the false positive, which refers to misclassified non-BAL QSOs as BALQSOs; $FN$ represents the false negative, which refers to misclassified BALQSOs as non-BAL QSOs.

\subsection{Comparison of Different Dimensionality Reduction Algorithms}
Before training the Random Forest and XGBoost algorithms, we employed dimensionality reduction techniques including PCA, LLE, ISOMAP, t-SNE, and a combination of PCA and t-SNE (PCA+t-SNE). Subsequently, the models were trained using 10-fold cross-validation. For PCA, we extracted the first 100 principal components to serve as input for the Random Forest and XGBoost algorithms. Regarding LLE and ISOMAP, we reduced the data to dimensions of 2, 10, and 100, respectively. In the case of t-SNE, we initially applied the technique to the entire dataset with default parameters, resulting in dimension reductions to two and three dimensions. Additionally, PCA was conducted on the dataset to reduce it to 100 dimensions before applying t-SNE with default parameters to obtain two and three-dimensional representations. These dimensionality-reduced datasets were subsequently used for model training. The accuracy of each model is summarized in Table \ref{table:PCA}.

\begin{longtable}{cccc}
    \caption{Classification results of PCA, LLE, ISOMAP, t-SNE and PCA+t-SNE.} \\
    \label{table:PCA}
    \hline 
    \hline
    \multicolumn{1}{c}{Reduction Method} & \multicolumn{1}{c}{No. of Dimensions} & \multicolumn{1}{c}{ML Model} & \multicolumn{1}{c}{Accuracy (\%)}  \\
    \hline 
    \endfirsthead
    
    \hline 
    \multicolumn{1}{c}{Reduction Method} & \multicolumn{1}{c}{No. of Dimensions} & \multicolumn{1}{c}{ML Model} & \multicolumn{1}{c}{Accuracy (\%)}  \\
    \hline \hline 
    \endhead
    
    \hline
    \hline
    \endlastfoot

    PCA & 100 & Random Forest & \bf{96.37} \\
    PCA & 100 & XGBoost & \bf{97.33} \\
    LLE & 2 & Random Forest & 76.78 \\
    LLE & 2 & XGBoost & 76.84 \\
    LLE & 10 & Random Forest & 92.39 \\
    LLE & 10 & XGBoost & 92.45 \\
    LLE & 100 & Random Forest &  94.21 \\
    LLE & 100 & XGBoost & 94.02 \\
    ISOMAP & 2 & Random Forest & 70.44 \\
    ISOMAP & 2 & XGBoost & 71.15 \\
    ISOMAP & 10 & Random Forest & 94.17 \\
    ISOMAP & 10 & XGBoost & 94.23 \\
    ISOMAP & 100 & Random Forest &  94.77 \\
    ISOMAP & 100 & XGBoost & 94.88 \\
    t-SNE & 2 & Random Forest & 92.82 \\
    t-SNE & 2 & XGBoost & 93.13 \\
    t-SNE & 3 & Random Forest & 93.51 \\
    t-SNE & 3 & XGBoost & 93.54 \\
    PCA+t-SNE & 100, 2 & Random Forest & 92.50 \\
    PCA+t-SNE & 100, 2 & XGBoost & 92.49 \\
    PCA+t-SNE & 100, 3 & Random Forest & 93.22 \\
    PCA+t-SNE & 100, 3 & XGBoost & 93.44 \\    

\end{longtable}

From Table \ref{table:PCA}, it is evident that PCA outperformed the other dimensionality reduction methods. The Random Forest model achieved an accuracy of 96.37\%, while the XGBoost model achieved an even higher accuracy of 97.33\%. The lowest accuracy was observed when reducing the data to 2 dimensions using LLE and ISOMAP. This result aligns with the visualization of the data in Section \ref{subsec:manifold_learning}, indicating the inability of the the two methods to effectively distinguish between the two types of QSOs. This underscores the superiority of PCA in preserving essential information for accurate classification.

Upon applying t-SNE, it was observed that the accuracy is marginally higher in three dimensions compared to two dimensions. The preceding application of PCA before t-SNE exhibited minimal impact on the accuracy. However, it is noteworthy that the accuracy of t-SNE dimensionality reduction models is markedly lower than PCA. Consequently, among our employed dimensionality reduction methods, PCA demonstrates the most efficacy in facilitating accurate classification.

Upon comparing the results presented in Table \ref{table:PCA}, it becomes apparent that the classification accuracy achieved by the XGBoost algorithm generally tends to be slightly higher than that of the Random Forest algorithm. However, the disparity between the two is not considerable, with the maximum difference being approximately 1\%. Interestingly, in certain instances, the accuracy of the Random Forest model may even marginally surpass that of XGBoost, with the maximum difference being around 0.2\%. This suggests that while the XGBoost algorithm exhibits slightly superior performance in addressing this classification problem, the distinction between the two algorithms is not substantial.

\subsection{Perplexity Parameter in t-SNE}

\begin{figure}[ht!]
\centering
\includegraphics[width=0.75\linewidth]{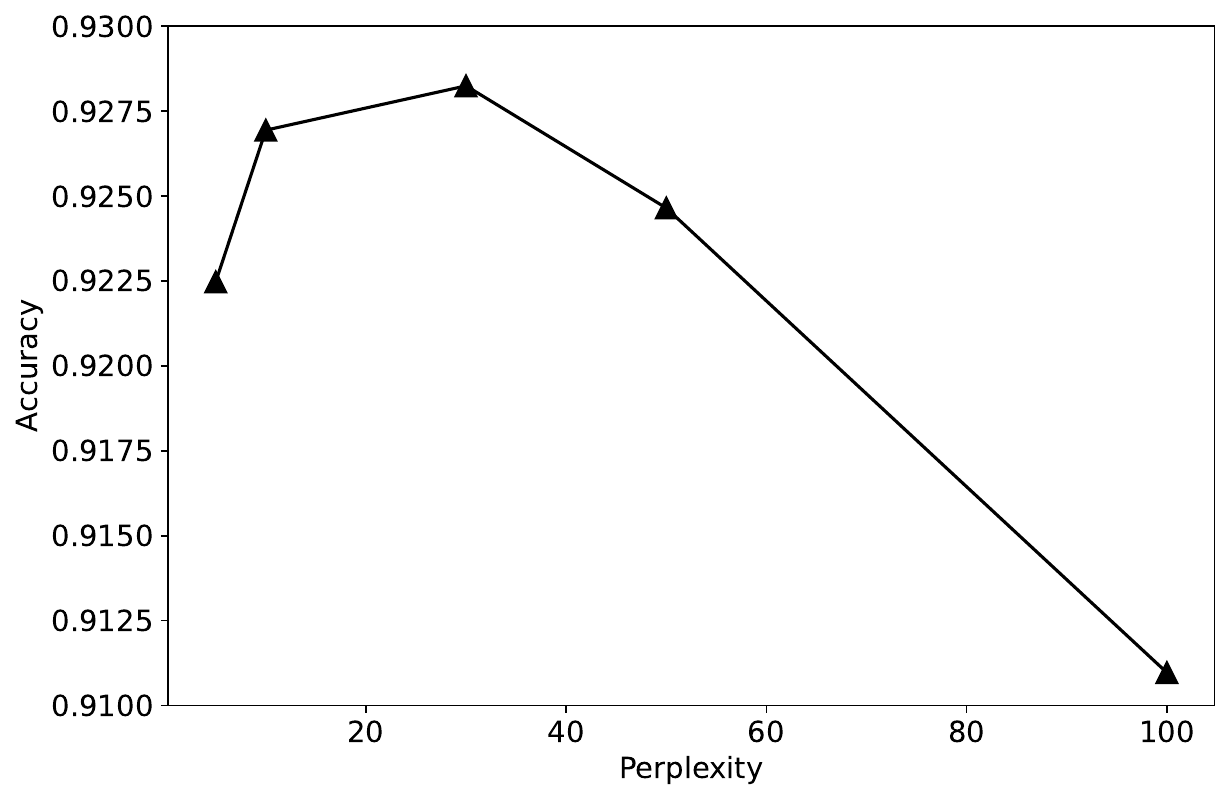}    
\caption{The classification accuracy of the Random Forest model as a function of ``perplexity" in t-SNE. The dataset is reduced to two dimensions by t-SNE. 
\label{fig:tsne_perlexity}}
\end{figure}

To assess the influence of different parameter values in the t-SNE dimensionality reduction algorithm, we conducted experiments by adjusting the ``perplexity" parameter to various values, ranging from 5 to 100, while reducing the dimensionality to two dimensions, as outlined in Section \ref{subsec:tsne}. The default value for perplexity is set to 30, and we explored alternative values including 5, 10, 30, 50, and 100. Subsequently, we input the dimensionality-reduced data generated by each perplexity value into the Random Forest model and evaluated their respective accuracies. The results, as depicted in Figure \ref{fig:tsne_perlexity}, reveal slight fluctuations in accuracy across different perplexity values. Notably, the highest accuracy is attained at the default value of 30, with values in close proximity to the default value also demonstrating commendable performance.

\subsection{Tuning Hyperparameters} \label{subsec:tune}

Hyperparameter tuning in machine learning models involves adjusting the values of their hyperparameters and observing the training results across different combinations of hyperparameters. Given the relatively limited impact of other hyperparameters (as discussed in Section \ref{subsec: intro_ML}), our primary focus was on tuning the ``n\_estimators" hyperparameter for the Random Forest model and the ``max\_depth" hyperparameter for the XGBoost model.

For the ``n\_estimators" hyperparameter in Random Forest, we experimented with various values such as 50, 100, 200, 300, 500, and 1000, with the default value set to 100. Regarding the ``max\_depth" hyperparameter in the XGBoost model, we tested a range of values including 3, 5, 6, 7, 9, 11, 13, and 15, with the default value set to 6.

\begin{figure*}[ht!]
\centering
\includegraphics[width=0.75\linewidth]{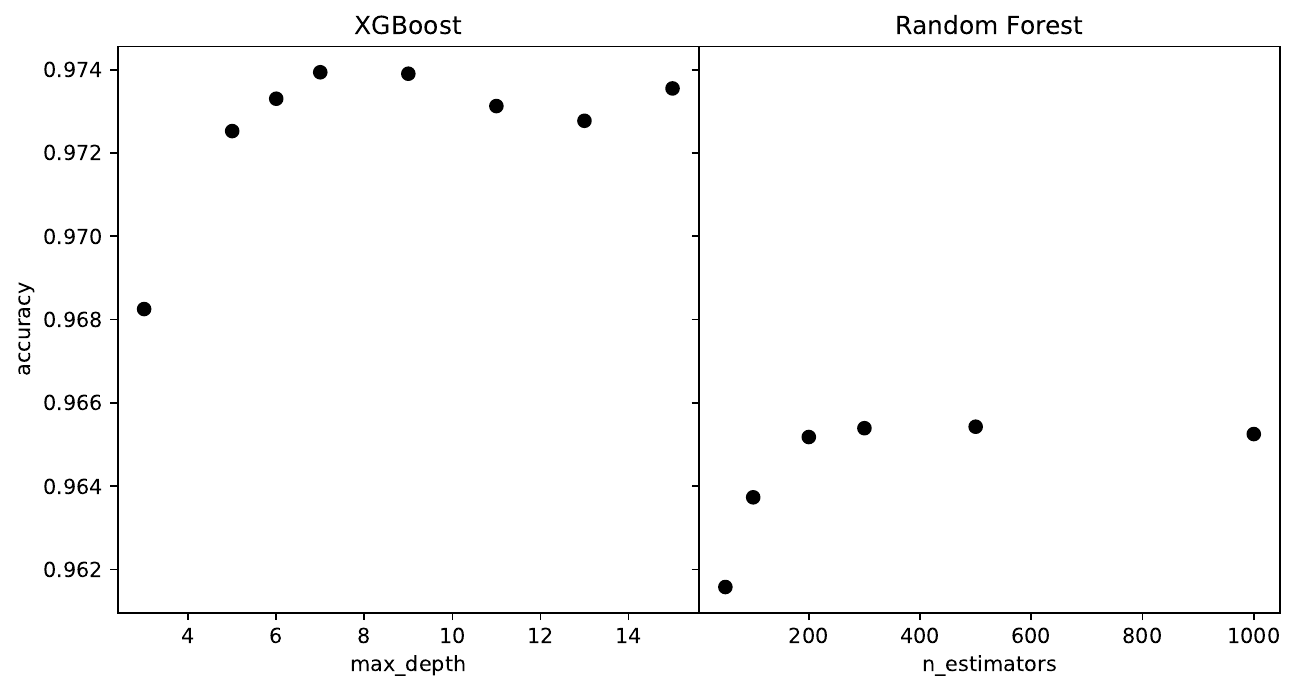}    
\caption{Accuracy as a function of ``max\_depth" in XGBoost (left panel) and ``n\_estimators" in Random Forest (right panel) when the first 100 principal components are taken as input. 
\label{fig:tune_parameter}}
\end{figure*}

The results, as illustrated in Figure \ref{fig:tune_parameter}, demonstrate that values around the default value of 6 for the ``max\_depth" hyperparameter in the XGBoost model generally yield satisfactory performance. The highest accuracy was achieved with $\mathrm{max\_depth}=7$, reaching 97.39\%, slightly surpassing the default value's accuracy of 97.33\%. When max\_depth is too small, the model lacks complexity and fails to effectively fit the data, resulting in lower accuracy. Conversely, excessively large values of max\_depth can lead to overfitting and a decrease in accuracy.

Regarding the ``n\_estimators" hyperparameter in the Random Forest model, increasing its value enhances accuracy due to the inclusion of more decision trees. However, larger values necessitate significantly more computational time and only yield marginal improvements in classification performance.

For this specific classification task, our experimentation with hyperparameter tuning revealed that neither XGBoost nor Random Forest demonstrated a substantial improvement in accuracy. Most of the hyperparameters exhibited negligible impact on the results, or they had already reached optimal accuracy levels near their default values, as observed with the max\_depth parameter in XGBoost. This indicates that adjusting the hyperparameters may not result in significant performance gains for this task.

\subsection{The Best Model}

Determining the optimal number of principal components to retain in PCA often involves a degree of subjectivity. To mitigate this inherent uncertainty, we opted to retain a larger number of principal components, ensuring that the cumulative variance surpassed the critical threshold of 80\%. This approach minimizes subjectivity and guarantees that the retained principal components faithfully capture the essential information embedded within the original spectrum.

Accordingly, we preserved a total of 448 principal components, collectively contributing to an accumulated variance of 80\% throughout the PCA analysis of the dataset. This meticulous selection process ensures that a significant portion of the original data's information content is preserved, laying a robust foundation for subsequent analyses and model training.

We utilized the retained 448 principal components to train both XGBoost and Random Forest models for the classification task, as detailed in Section \ref{subsec:training method}. Employing default hyperparameters, the XGBoost model attained an accuracy of 97.57\%, whereas the Random Forest model achieved an accuracy of 95.82\%. Comparing these results to those obtained by reducing the dimensionality to 100 dimensions using PCA, we observe that the accuracy of the XGBoost model slightly increased from 97.33\% to 97.57\%. Conversely, the accuracy of the Random Forest model decreased from 96.37\% to 95.82\%. These findings suggest that XGBoost can effectively leverage a higher number of dimensions as input to yield a more accurate model. However, it appears that the Random Forest model may be susceptible to overfitting and exhibit diminished performance with increased dimensions.

These insights underscore the importance of selecting an appropriate machine learning algorithm and dimensionality reduction strategy tailored to the specific characteristics of the dataset and classification task at hand. Moreover, they highlight the nuanced interplay between model complexity, dimensionality, and performance, warranting careful consideration in model selection and parameter tuning to optimize classification accuracy.

Subsequently, akin to the approach outlined in Section \ref{subsec:tune}, we conducted hyperparameter tuning, albeit observing only marginal improvements in accuracy. In our pursuit of further enhancing performance, we delved into hyperparameter tuning, with a particular focus on optimizing the ``max\_depth" hyperparameter within the XGBoost model. Through meticulous experimentation, we identified the optimal setting for ``max\_depth" to be 5. Consequently, the model configured with ``max\_depth" set to 5 emerged as our top-performing model, boasting an impressive accuracy of 97.60\%. 

\begin{figure*}[ht!]
\centering
\includegraphics[width=0.9\linewidth]{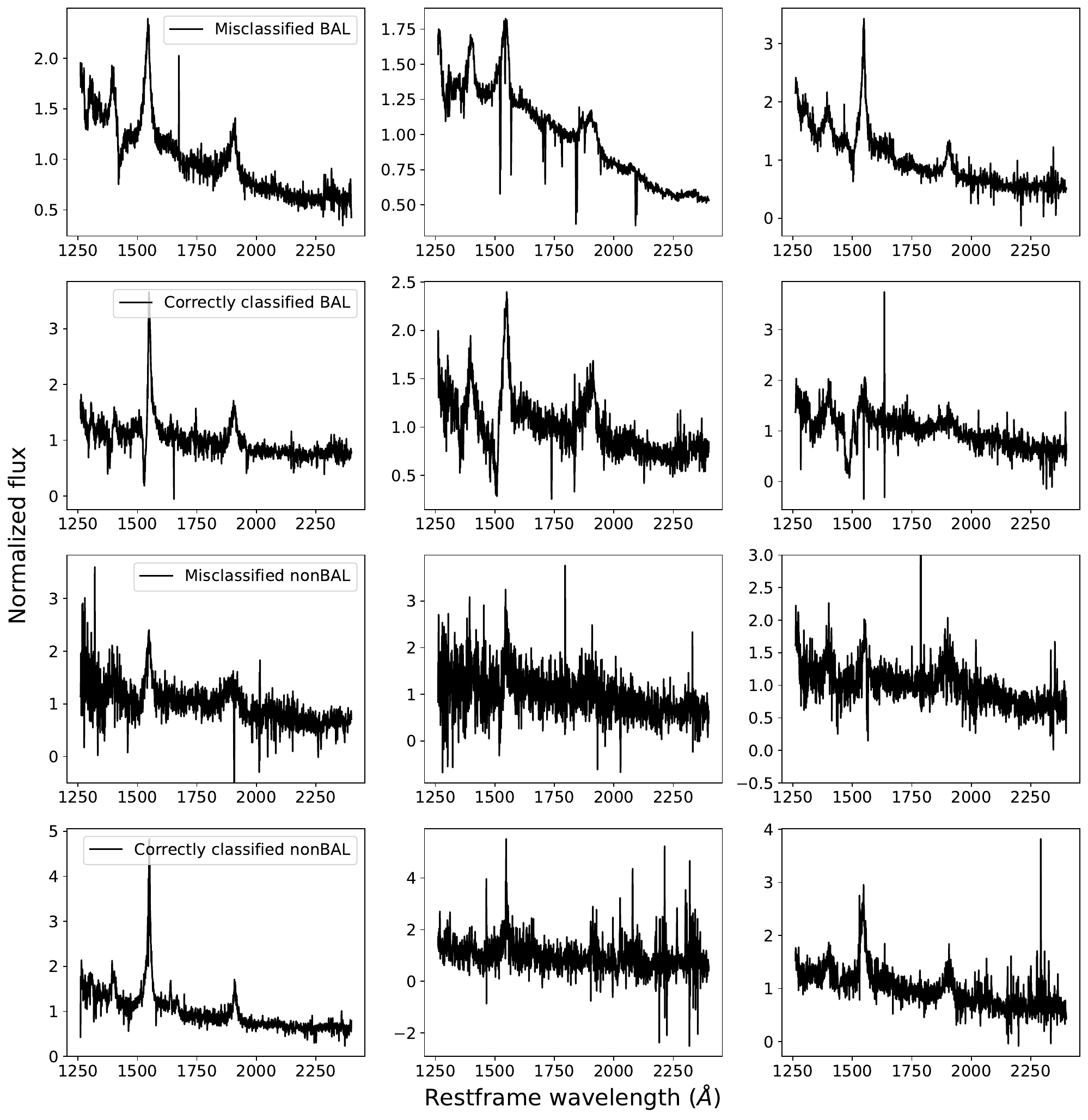}    
\caption{Some spectra of BALQSOs and non-BAL QSOs in the outer test sample. First row: BALQSO spectra that are misclassified as non-BAL QSOs; Second row: BALQSOs that are correctly classified; Third row: non-BAL QSOs that are misclassified as BALQSOs, and fourth row: non-BAL QSOs that are correctly classified.
\label{fig:classified_result}}
\end{figure*}

We evaluated the performance of our best-performing model by the outer test sample to assess generalization. The accuracy is 96.92\% for the outer test sample, which is lower than that obtained from cross-validation. The confusion matrix is depicted in Table~\ref{table: Confusion matrix}. For BALQSOs, Recall, Precision and F1-Score are 97.65\%, 96.24\% and 96.94\%, respectively, while for non-BAL QSOs, they are 96.19\%, 97.61\% and 96.90\%, respectively. Figure \ref{fig:classified_result} shows some examples about the misclassified and correctly classified spectra. The misclassification of BAL spectra may be attributed to the absence of continuum spectrum subtraction, which decreases with increasing wavelength. Additionally, the misclassification of non-BAL spectra could be due to low signal-to-noise ratio and slight absorption features.

\begin{table}
\caption{Confusion matrix for applying the best model on the outer test sample.}
\label{table: Confusion matrix}
\begin{tabular}{cccc}
\toprule
& Recall  & Precision & F1-Score\\
\midrule
BALQSOs   &97.65\%  & 96.24\% &96.94\%  \\
Non-BAL QSOs&96.19\%  & 97.61\% &96.90\% \\ 
\bottomrule
\end{tabular}
\end{table}

\section{Discussion} \label{sec:discussion}

\cite{Guo2019} classified quasar spectra in SDSS DR14 using Convolutional Neural Networks (CNNs). They divided quasars into BAL quasars and non-BAL quasars, achieving an accuracy of 98\%. \cite{Busca2018} developed QuasarNET, a deep CNN trained on quasar spectra from BOSS, which could distinguish BAL quasars with 98.0\% accuracy and non-BAL quasars with 97.0\% accuracy.

While our study yielded comparable accuracy to previous research, it's crucial to acknowledge the inherent limitations in directly comparing results due to disparities in datasets and machine learning methodologies employed. Our research endeavors to provide a nuanced exploration of various combinations of dimensionality reduction techniques and machine learning models, alongside fine-tuning their hyperparameters, to identify the optimal solution tailored to our specific classification task.

By systematically investigating four distinct dimensionality reduction methods and two classification machine learning models, we aim to unearth the most effective approach for spectral classification of quasars. Moreover, the outcomes derived from our comprehensive analysis can serve as a benchmark for validation and comparison with traditional manual classification methods.

This holistic approach not only advances our understanding of quasar spectral classification but also furnishes invaluable insights for enhancing classification accuracy in future endeavors. By elucidating the efficacy of different combinations of methodologies, our research contributes to the ongoing refinement of astronomical classification techniques, thereby bolstering the scientific rigor and precision of spectral analysis in astrophysical research.

\section{Conclusion} \label{sec: conclusion}
In this study, we explored multiple dimensionality reduction methods, including PCA, t-SNE, LLE, and ISOMAP, in combination with the Random Forest and XGBoost machine learning models for classification tasks. Through the application and comparison of these methods, we have drawn the following conclusions:

\begin{enumerate}
    \item PCA emerges as the most effective dimensionality reduction method for the QSO classification task. When PCA is applied to reduce the dimensionality of the input data for machine learning models, it yields the highest classification accuracy. With default parameters, the Random Forest model achieves 96.37\% accuracy, while the XGBoost model achieves 97.33\%. 
    \item LLE and ISOMAP, as manifold learning methods, exhibit similar classification results. When reducing the data to two dimensions, neither method effectively separates the two classes of QSOs, both visually and in terms of classification accuracy. Dimensionality reduction to 10 dimensions is necessary to achieve accuracy over 92\%, and even when reducing the data to 100 dimensions, the highest accuracy achieved is only 94.88\%, still lower than PCA results.
    \item The t-SNE algorithm, regardless of whether the data is first reduced using PCA, yields similar final classification results after dimensionality reduction. Accuracy slightly improves when reducing the data to three dimensions compared to two dimensions. However, the highest accuracy achieved is only 93.54\%, which even falls below the results obtained from LLE and ISOMAP dimensionality reduction.  
    \item In this classification problem, XGBoost generally outperforms Random Forest in accuracy, although the difference between the two is not significant.
    \item Adjusting the ``perplexity" parameter in the t-SNE algorithm results in slight differences in classification accuracy, with the highest accuracy obtained with values close to the default value.
    \item Hyperparameter tuning for both XGBoost and Random Forest models does not significantly enhance model accuracy. Most hyperparameters have minimal impact on results, or optimal accuracy is already attained with values close to the default settings.
\end{enumerate}

The pinnacle of accuracy is achieved by initially implementing PCA for dimensionality reduction, followed by training an XGBoost model for QSO classification. To ensure the fidelity of information preservation during PCA, we carefully select and retain 448 principal components, capturing just over 80\% of the cumulative variance from the original dataset. These principal components serve as input for training the XGBoost model, where hyperparameters are meticulously tuned. Notably, the optimal ``max\_depth" parameter is determined to be 5, resulting in an exceptional accuracy of 97.60\%. Additionally, the model achieves an accuracy of 96.92\% on the outer test sample.

The amalgamation of PCA dimensionality reduction with the XGBoost model not only enhances accuracy but also offers scalability and interpretability. Furthermore, by retaining a greater number of principal components and fine-tuning hyperparameters, we observe incremental improvements in model performance.

Our research findings hold broader implications beyond the realm of quasar classification, serving as a valuable reference for other astronomical spectroscopic classification endeavors. Researchers engaged in spectroscopic surveys, such as LAMOST, DESI, and analogous initiatives, can leverage our methodology to optimize their approach, select appropriate dimensionality reduction algorithms, machine learning models, and fine-tune relevant parameters, thereby facilitating advancements in astronomical classification techniques. Through the amalgamation of computational prowess with astronomical inquiry, machine learning promises to unravel the intricate tapestry of the universe, heralding a new era of discovery and understanding.

\section*{Data Availability}
The data used in this article are available in the SDSS website, at \url{https://dr18.sdss.org/}.

\begin{ack}
We are very grateful to the reviewer for his valuable feedback and insightful comments. The study was funded by the National Natural Science Foundation of China under grants Nos. 12273076 and 12133001, the China Manned Space Project with science research grant Nos. CMS-CSST-2021-A04 and CMS-CSST-2021-A06. We acknowledge SDSS databases. Funding for the Sloan Digital Sky Survey IV has been provided by the Alfred P. Sloan Foundation, the U.S. Department of Energy Office of Science, and the Participating Institutions. SDSS-IV acknowledges support and resources from the Center for High-Performance Computing at the University of Utah. The SDSS website is www.sdss.org. SDSS-IV is managed by the Astrophysical Research Consortium for the Participating Institutions of the SDSS Collaboration including the Brazilian Participation Group, the Carnegie Institution for Science, Carnegie Mellon University, the Chilean Participation Group, the French Participation Group, Harvard-Smithsonian Center for Astrophysics, Instituto de Astrof\'isica de Canarias, The Johns Hopkins University, Kavli Institute for the Physics and Mathematics of the Universe (IPMU) /University of Tokyo, Lawrence Berkeley National Laboratory, Leibniz Institut f\"ur Astrophysik Potsdam (AIP), Max-Planck-Institut f\"ur Astronomie (MPIA Heidelberg), Max-Planck-Institut f\"ur Astrophysik (MPA Garching), Max-Planck-Institut f\"ur Extraterrestrische Physik (MPE), National Astronomical Observatories of China, New Mexico State University, New York University, University of Notre Dame, Observat\'ario Nacional / MCTI, The Ohio State University, Pennsylvania State University, Shanghai Astronomical Observatory, United Kingdom Participation Group, Universidad Nacional Aut\'onoma de M\'exico, University of Arizona, University of Colorado Boulder, University of Oxford, University of Portsmouth, University of Utah, University of Virginia, University of Washington, University of Wisconsin, Vanderbilt University, and Yale University.

\end{ack}

\bibliography{example}{}
\bibliographystyle{aasjournal}



\end{document}